\documentclass[10pt]{article}
\begin{document}
\title{ Solutions of time-dependent Schr\"{o}dinger equations for model non-Hermitian quantum mechanical systems }
\author{ Brian L Burrows\\   Emeritus Professor, University of Staffordshire\\ College Road,University Quarter, Stoke-on-Trent, ST4 2DE UK  \\ \small{e-mail: brian.burrows2@btopenworld.com} }
 \maketitle
\section*{Abstract}
The time-dependent Schr\"{o}dinger equation is solved for two model problems for a non-Hermitian quantum system.A simple matrix model system is used to examine two critical problems for these systems: complex and  non-observable energies and situations where the matrix is defective. In addition the stationary states  for  infinite dimensional model system, which is confined in space , is examined and it is shown that the pattern of eigenvalue energies differs from when the system is unconfined.
\section{Introduction}
In recent years there has been increased interest in non-Hermitian  quantum systems.Most of these have treated the problems  as a time-independent systems. The main interest has been in stationary states  but these are just particular solutions of the time-dependent Schr\"{o}dinger equation. In this paper the  time-dependent theory for non-Hermitian operators  is considered and two examples are presented, one of which is confined in space. \newline
The first is a 2-dimensional model where the matrix operator is non-Hermitian. This gives rise to two problems that are not present when the matrix is Hermitian. The eigenvalues of the operator may be complex and hence non-observable. To construct an observable quantity, related to the energy eigenvalue,  a time -dependent transform is used , which is a simple example of the earlier work described in  in [1-3].The other problem is that in particular cases the eigenvalues and eigenvectors may coalesce so that the matrix is a defective and only has one eigenvector. The stationary states are particular solutions of the time-dependent equation and in general can be used as a basis for all solutions. But for the defective matrix we can still find solutions with a basis of the eigenvector and an arbitrary vector independent of the eigenvector, conveniently a vector orthogonal to the eigenvector. Thus contrary to what might be thought, the defective matrix case leads to a solution in the same  way. However for the defective matrix there is only one stationary state. These observations apply to other systems where we may use the stationary states as a basis, but in some cases the basis needs to be augmented by arbitrary independent functions.\newline
The second example  is a one variable problem where the underlying Hilbert space  is infinite dimensional and we consider  the calculation of the stationary states. The example is well known for an infinite range of the variable but here a confined system is considered  so that wave function is confined in space. The space confinement leads to a different distribution of real and complex eigenvalues which differs from the unconfined problem. This is examined further in an Appendix.\newline
Both of the examples considered are Hamiltonians that are PT symmetric which implies that either the corresponding eigenvalues  are either real or complex conjugate pairs. There is an extensive literature on non-Hermitian problems and in particular on PT symmetric systems and a selection is given, which includes review articles with a wider selection of references [4-10 ].More recent work  has included the theory of time -dependent transformation which we have used in the next section [1].
\section{ A simple example}
We consider the time-dependent Schr\"{o}dinger equation 
\begin{equation}
i\frac{\partial \psi}{\partial t} = H\psi, \quad \psi(0) =\psi_{0} 
\end{equation}
where the operator  $ H $ is the non-Hermitian two dimensional matrix
\[H=
\left(
\begin{array}{cc}
1-i  &  \mu   \\
 \mu &1+ i     
\end{array}
\right)
\]
 with real $ \mu >0 $ and $\psi_{0} $ is a given 2-dimensional vector. The eigenvalues of $ H$ are $ \lambda_{1}=1+\sqrt{\mu^2-1}, \lambda_{2}=1- \sqrt{\mu^2-1}$. So for varying $ \mu $ we can obtain two real eigenvalues, or two imaginary eigenvalues and there is a special case where both eigenvalues are 1.The special case  will be considered separately but in general the eigenvectors corresponding to $ \lambda_{1},\lambda_{2} $ are
\[
\mathbf{u}_{1} =\left(
\begin{array}{c}
  \mu   \\
  \lambda_{1}-1 +i   
\end{array}
\right),
\mathbf{u}_{2}=\left(
\begin{array}{c}
  \mu   \\
  \lambda_{2} -1+i   
\end{array}
\right)
\]
We note that for $\mu^2=1$ we have the special case of a repeated eigenvalue but also the eigenvectors coalesce,  which we examine later, but in general for some choice of constants $ A, B$  we have $\psi_{0} = A \mathbf{u}_{1} + B\mathbf{u}_{2}$  and the solution may therefore be expressed in terms of a basis :\begin{equation}
\psi(t) =A\psi_{1}(t) + B\psi_{2}(t)= Aexp(-i\lambda_{1}t) \mathbf{u}_{1}+ Bexp(-i\lambda_{2}t)\mathbf{u}_{1}
\end{equation}
For the particular special cases where $A=0 $ or $B=0 $ we have $ |\psi(t)|^2 $ independent of time; that is stationary states.

We can transform the basis  using the  transformation to $\psi=U\hat{\psi }$ where $U $ is the constant matrix whose columns are $ \mathbf{u}_{1},\mathbf{u}_{2}$  so that $H$ transforms to the diagonal matrix $ h=U^{-1}HU $ and the corresponding  time-dependent equations for the stationary states are then conveniently written in the form
\begin{equation}
i\frac{\partial \hat{\psi}}{\partial t} =i\frac{\partial}{\partial t} 
\left(
\begin{array}{c}
 \psi_{1}    \\
 \psi_{2}    
\end{array}
\right)= 
\left(
\begin{array}{cc}
 \lambda_{1} &  0  \\
  0   & \lambda_{2}  
\end{array}
\right)\left(
\begin{array}{c}
 \psi_{1}    \\
 \psi_{2}    
\end{array}\right)=h\left(
\begin{array}{c}
 \psi_{1}    \\
 \psi_{2}    
\end{array}
\right)
\end{equation}
For $ \mu^2>1 $ we obtain two  real , observable energies $1 \pm\sqrt{\mu^2-1}$ but for $ \mu^2<1 $ these energies are $1 \pm i\sqrt{1-\mu^2}$ which are unobservable. In order to obtain observable quantities we may use an \emph{additional time-dependent transformation},  of the form

\begin{equation}
\left(
\begin{array}{c}
 \psi_{1}    \\
 \psi_{2}    
\end{array}\right)= \eta(t) \phi(t) = 
\left(
\begin{array}{cc}
 r(t) &   0  \\
 0  &   s(t)   
\end{array}
\right)\phi(t)
\end{equation}
 with $r(t)>0, s(t)>0$, so that
 \begin{equation}
i\frac{\partial}{\partial t} 
\left(
\begin{array}{c}
 \phi_{1}    \\
 \phi_{2}    
\end{array}
\right)= (\eta^{-1}h \eta -i\eta^{-1}\frac{d\eta}{dt} )\phi
\end{equation}
where
\[ \eta^{-1} =\left( \begin{array}{cc}
\frac{1}{ r(t)} &   0  \\
 0  &   \frac{1}{s(t) }  
\end{array}
\right), \frac{d\eta}{dt} =\left(
\begin{array}{cc}
 \frac{dr}{dt}&   0  \\
 0  &   \frac{ds}{dt}   
\end{array}
\right) \]
 
 so that
 
\begin{equation}
i\frac{\partial}{\partial t} 
\left(
\begin{array}{c}
 \phi_{1}    \\
 \phi_{2}    
\end{array}
\right)= 
\left(
\begin{array}{cc}
 \lambda_{1} &  0  \\
  0   & \lambda_{2}  
\end{array}
\right)\left(
\begin{array}{c}
 \phi_{1}    \\
 \phi_{2}    
\end{array}\right)-\left(
\begin{array}{cc}
\frac{i}{r} \frac{dr}{dt}&  0  \\
  0   & \frac{i}{s}\frac{ds}{dt} 
\end{array}
\right)\left(
\begin{array}{c}
 \phi_{1}    \\
 \phi_{2}    
\end{array}\right)
\end{equation}
 and this time-dependent transformation does not preserve the eigenvalues. For real eigenvalues we may choose  $ r=s=1$ and the eigenvalues are unchanged, but for complex eigenvalues we choose 
 \begin{equation}
  \frac{dr}{dt} = r \sqrt{\mu^2-1}  \Rightarrow  r =  \frac{exp( t\sqrt{\mu^2-1})}{exp( \tau \sqrt{\mu^2-1} )}<1,\quad \tau >0
\end{equation}
so that we remove imaginary part of $ \lambda_{1} $. A similar choice  $ s =  \frac{exp(- t\sqrt{\mu^2-1})}{exp(- \tau \sqrt{\mu^2-1} )} $ removes the imaginary part in $ \lambda_{2} $. This procedure is essentially a transformation from Hilbert space with inner product $ < > $ to a new space where  the new inner product is\begin{equation}
(\phi_{1} |\phi_{2}) = < \phi_{1} | (\eta)^{\dagger} \eta \phi_{2}>
\end{equation}
The new operator, $h_{1} $  is Hermitian in this space and the eigenvalues are real and observable since they are simply the real parts of the original eigenvalues of $ H $. Merzbacher [11] , in the discussion of stationary states, argues that in nature  there are physically stable states , not necessarily mathematically precise stationary states and so that this form of analysis is not unphysical. In $ 0 \leq t \leq \tau $ we have $ r>1$ and $ s>1 $ so that the transformation is well defined. However since $ \tau $ is arbitrarily large the transformation may be applied for arbitrary $t>0 $.

If we now consider the special case  where $ \mu =1 $ there is only one eigenstate of $ H $ and hence only one stationary solution. But we can still use an analogous basis as in (2)  where the second function is an \emph{ arbitrary vector } which is orthogonal to the eigenvector and hence independent of it. Specifically we choose the second vector to be the eigenvector of $ H^{\dagger} $  corresponding to this shared eigenvalue. Note that for the standard analysis for $\mathbf{u}_{1}$ and $\mathbf{u}_{2} $ being the eigenvectors of a general $ H $ and $ H^{\dagger}$ respectively then $ \mathbf{u}_{1}^{\dagger}\mathbf{u}_{2} =0 $ if the corresponding eigenvalues are different  but arbitrary if the eigenvalues are identical.However for the situation
\begin{equation}
H \mathbf{u}_{1} = \lambda \mathbf{u}_{1},H^{\dagger} \mathbf{u}_{2}=\lambda \mathbf{u}_{2},H\mathbf{v}_{1}= (\lambda + \epsilon)\mathbf{v}_{1}, H^{\dagger}\mathbf{v}_{2}= ( \lambda + \epsilon)\mathbf{v}_{2}
\end{equation}
we have $ \mathbf{v}_{2}^{\dagger} \mathbf{u}_{1} =0 $ and as $ \epsilon \rightarrow 0 $ we obtain $ \mathbf{u}_{2}^{\dagger} \mathbf{u}_{1} =0 $.
So  that  in the case where the eigenvalues coalesce we  can choose $ \mathbf{u}_{2} $ as the arbitrary independent vector. Thus
\begin{equation}
H \mathbf{u}_{1} = \left(
\begin{array}{cc}
 1-i &  1  \\
  1   & 1+i  
\end{array}
\right)\left(
\begin{array}{c}
 1   \\
 i   
\end{array}\right)= \mathbf{u}_{1}, \quad \mathbf{u}_{1} ^{\dagger}\mathbf{u}_{1}=2
\end{equation}
and for the choice  $ \mathbf{u}_{2} = ( 1,-i)^{T} $ we have
\begin{equation}
H\mathbf{u}_{2} = \left(
\begin{array}{c}
 1-2i   \\
 2-i   
\end{array}\right), \quad \mathbf{u}_{2} ^{\dagger}\mathbf{u}_{2}=2
\end{equation}
Furthermore $ \mathbf{u}_{2} ^{\dagger}H\mathbf{u}_{1}=0,\mathbf{u}_{1} ^{\dagger}H\mathbf{u}_{2}=-4i $  and $ \mathbf{u}_{2} ^{\dagger}H\mathbf{u}_{2}=2$. So that writing $\psi = a \mathbf{u}_{1} + b\mathbf{u}_{2}$  in (1) we obtain the differential equations:\begin{equation}
\frac{da}{dt} = -ia -2b, \frac{db}{dt} = -ib 
\end{equation}
with the solution \begin{equation}
a= K_{2}exp(-it) -2 K_{1} texp(-it), b=K_{1}exp(-it)
\end{equation}
The arbitrary constants, $ K_{1},K_{2} $ can be chosen to match  any given $ \psi_{0} $.Choosing $ K_{1} =0 $ gives the stationary state.The analysis for $ \mu=-1 $ is analogous. The Schr\"{o}dinger equation is a time-dependent equation and this analysis illustrates we can solve the equation for all $\mu $ and the solution in this special case is equally valid, mathematically and physically,  to any other case. The only difference is the number of stationary state solutions.The stationary states can be specified by
\begin{equation}
i\frac{\partial \psi}{\partial t} = H\psi, \psi(0) =\psi_{0}, H\psi_{0} = E\psi_{0} 
\end{equation}
Of course any of  the sets of basis functions constructed here can be used  any matrix of the same form   and it it is not necessary to restrict the basis to be eigenvectors  of the matrix.Given a fixed basis we then have a set of equations that can be applied for all  $ \mu $. The limit of the eigenvalues as $ \mu \rightarrow 1 $ is smooth  but there are singularities in the variation of the  eigenvectors in this limit.Thus for $ \mu \approx 1 $  it may be more convenient to write $ H$ in the form  
\begin{equation}H=
\left(
\begin{array}{cc}
1-i  &  1   \\
 1 &1+ i     
\end{array}
\right)+ (\mu-1) \left(
\begin{array}{cc}
0  &  1   \\
 1 &0     
\end{array}
\right)
\end{equation}
and to use the basis $ \mathbf{u}_{1} = ( 1,i)^{T} , \mathbf{u}_{2} = ( 1,-i)^{T} $.

\section{An example of an infinite dimensional Space}
A particular example of an infinite system  in a \emph{finite space } interval  is 
\begin{equation}
i\frac{\partial \Psi}{\partial t}  =H\Psi =-\frac{\partial^2 \Psi}{\partial x^2} + i\mu x\Psi, \quad -T/2<x<T/2, \Psi (\pm T/2) =0
\end{equation}
and to analyse the stability of the system  we calculate the stationary states so that the boundary conditions are  $ \Psi(0) = \psi, H\psi =E\psi $ for some $E $ which may be real or complex. 
For the given operator $ H $ to be defined precisely we need the domain and the consequent range  the the space  considered is the subset of the Hilbert space generated by the following set of orthogonal functions
\begin{equation}  \{sin( \frac{2m \pi x}{T}) , cos(\frac{(2n-1)\pi x}{T}), \quad n,m =1..\infty\} \end{equation} with the  inner product for arbitrary functions $ \chi_{1},\chi_{2} $
\begin{equation}
< \chi_{1} |\chi_{2}> =\frac{1}{T} \int_{-T/2}^{T/2} \chi_{1}(x)^{*}\chi_{2}(x) dx
\end{equation}

The Hilbert space would be completed by $ \chi= 1 $ but this function is not in the domain of the operator.This model is essentially different from using the Hermite functions in  $ -\infty < x < \infty$ since the derivatives of the basis functions are  not zero at the end points.However the probability density function is zero at the end points. 
First of all consider some numerical examples using finite matrices where we use $ N $ even functions and N odd functions. The values given are the real parts of the eigenvalues so that the repeated values are the complex conjugate pairs.  For \emph{fixed} $ T $ as we increase $ N$  we get an increase in the number of  real eigenvalues but for an increase in $T$  we get an increase in the number of complex eigenvalues.. For $ T=12 $ as shown in table 1  there are only 6 eigenvalues that form complex conjugate pairs. If we now decrease $ T$ we can reach a stage where all eigenvalues are real and this is illustrated in table 2 where convergence is indicated as $ N $ increases. In the last 4 entries we see the smooth but rapid change from complex pairs to real eigenvalues. For the complex pair the imaginary part is $ \pm i (0.0886971) $ whereas for the two final real states the imaginary part is zero(  The numerical roundoff being $ 10^{-13} $). \newline The results in table 3 illustrate the change in the eigenvalues as $T $ and  $ \mu $ are increased. The pattern is the same but the number of complex eigenvalues increases.The pattern displayed is general.
\begin{table}[htp]
\caption{T=12,Matrix=2N x 2N,  $\mu=1$}
\begin{center}
\begin{tabular}{|c|c|c|c|}
N=10&N=12&N=20& N=40\\
1.16905477&1.16905391&1.16905371&1.16905371\\
1.16905477&1.16905391&1.16905371&1.16905371\\
2.04397669&2.04397513&2.04397477&2.04397477\\
2.04397669&2.04397513&2.04397477&2.04397477\\
2.76026162&2.76025943&2.76025892&2.76025891\\
2.76026162&2.76025943&2.76025892&2.76025891\\
3.32774482&3.32773931&3.32773734&3.32773734\\
3.72377774&3.72378086&3.72378299&3.72378306\\
5.05683219&5.05683027&5.05683108&5.05683113\\
6.44320495&6.44320804&6.44321054&6.44321061
\end{tabular}
\end{center}
\label{default}
\end{table}%

\begin{table}[htp]
\caption{Matrix=2N x 2N, $\mu=1$}
\begin{center}
\begin{tabular}{|c|c|c|c|}
T& N&State& E\\
6&2&1&1.16754817\\
6&2&2&1.16754817\\
6&2&3&2.44324110\\
6&20&1&1.16277827\\
6&20&2&1.16277827\\
6&20&3&2.25704001\\
4.63&40&1&1.13291267\\
4.63&40&2&1.13291267\\
4.6182&40&1&1.3269365\\
4.6182&40&2&1.3394567 
\end{tabular}
\end{center}
\label{default}
\end{table}%

\begin{table}[htp]
\caption{N=40, varying $T$ and $\mu$}
\begin{center}
\begin{tabular}{|c|c|c|}
$E(T=12,\mu=1) $ &$E(T=13,\mu=1$ & $E( T=12,\mu=1.5)$\\
1.16905371& 1.16905370&1.53189372\\
1.16905371& 1.16905370&1.53189372\\
2.04397477&2.04397478&2.67836460\\
2.04397477&2.04397478&2.67836460\\
2.76025892&2.76028890&3.61698875\\
2.76025892&2.76028890&3.61698875\\
3.32773734&3.38761029&4.44725786\\
3.72378306&3.38761029&4.44725786\\
5.05683113&4.10690531&5.03432252\\
6.44321061&5.28902379&5.96280806

\end{tabular}
\end{center}
\label{default}
\end{table}%

In order to explain this pattern consider  fixed $ T $ and large $ N$  and note that the matrix elements satisfy
\begin{equation}
h_{n,n} = \frac{n^2\pi^2}{T^2} = O(n^2/T^2), h_{2n-1,2n}=h_{2n,2n-1} = i\frac{1}{T} \int_{-T/2}^{T/2} x\omega_{2n-1}\omega_{2n}dx=O(T)
\end{equation}
where 
\begin{equation}
\omega_{2n-1} = cos(\frac{(2n-1)\pi}{T} , \omega_{2n}=sin(\frac{2n\pi x}{T} )
\end{equation}
and all other matrix elements are zero. For  $ n $ large the $n^{th} $ row is diagonally dominant ( since the sum of the moduli of the off diagonal elements is $O(nT) $ and this generally is a severe overestimate),  so that we have eigenvalues in the form $ h_{n,n} + \epsilon_{n} $ where $ \epsilon_{n}<<h_{n,n} $ . Furthermore $h_{n,n} $ increases rapidly with $n$ so repeated real parts do not occur and consequently we have real eigenvalues and not the complex conjugate pairs.As $T$ increases the number of complex pairs increases and  it is well known that in the infinite interval the  eigenvalues are not real. This is illustrated in the appendix where the space confinement is replace by an asymptotic form.
\section{Conclusion}
Here the time-dependent solutions of the Schr\"{o}dinger equation, particularly for stable stationary states, are examined using model problems. The results  from the simple matrix problem show that even for a defective matrix, solutions exist but the number of stationary states may be reduced. When the eigenvalues are complex a time-dependent transformation to a different Hilbert space with a time-dependent metric can be used, to find observable quantities that are simply the real parts of the complex eigenvalues. In the second, infinite dimensional problem it is shown that the operator domain and range need to be considered and a space confinement is imposed. The confinement gives rise to a different pattern of stationary state energies for which the higher states must be real. This is compared to the infinite state in the appendix.
\section{Appendix:The relationship between the finite and infinite interval}

In order to examine how these  problems are related to similar problems for an infinite interval ( when $ T \rightarrow \infty $) we consider an operator  in the Hilbert space of square integrable  functions in $ -\infty <x < \infty  $ so that $H=-D^2 +ix^m $, $m$ an odd integer ($D=\frac{d}{dx} $), where the domain is the space of continuous and differentiable functions, (including  piecewise continuous and differentiable continuous functions) so that the second derivative is finite  . For a more precise definition of the operator we need to define the asymptotic boundary conditions analogous to the wave function being zero at $ x=\pm T/2 $ in the finite interval. Essentially the asymptotic analysis replaces the confinement in a finite interval.There are many possibilities for the behaviour at $ \pm \infty$ but we require the wave function to be normalisable. One possibility is where  for  $ x>0 $, the wave function can be expressed in the form 
\begin{equation}
\psi =exp(-bx ^p)\psi_{1},   Re(b)>0 
\end{equation}
where $ \psi $ is continuous and differentiable throughout the region which implies that $ D^{2}\psi $ is finite  
and where  $ \psi_{1} $ is $ O(x^q) $  for some   $ q $ as $ x\rightarrow \infty $ Thus we have 
\begin{equation}
(-D^2+ ix^m)\psi =E \psi,  -\infty < x< \infty  \end{equation}
which implies that
\[ (-D^2\psi_{1} + bp(p-1) x^{p-2}\psi_{1} +2bpx^{p-1}D\psi_{1} -b^2p^2x^{2p-2}) \psi_{1} ) + \]
 \begin{equation} ix^m \psi_{1} - E\psi_{1} = 0
\end{equation}
Choosing
\begin{equation}
2p-2 = m , b^2p^2 =i
\end{equation}
eliminates the potential term $ ix^m $ and leads to $ p=(m+2)/2 $ and $ b= \frac{2}{m+2} \sqrt(i).$ This implies
\begin{equation}
b= \frac{2}{m+2} ( cos(\pi/4) + isin(\pi/4 ) = \frac{\sqrt{2}}{m+2} (1+i)
\end{equation}
since $ Re(b) >0 $.The equation for $ \psi_{1} $ becomes
\begin{equation}
(-D^2 + b\frac{m^2+2m}{4} x^{(m-2)/2} +(m+2)bx^{m/2}D -E)\psi_{1} =0
\end{equation}
From the assumptions of continuity on $ \psi_{1} $ we have that $D^2\psi_{1} $ remains finite as $ x\rightarrow 0 $ so that for equations (21) and (22) to be consistent we need $ m \geq  2 $.
 Now as $ x\rightarrow \infty $, from the earlier assumption, we may express $ \psi_{1} $ in the form 
\begin{equation}
\psi_{1} = x^q ( \sum_{n=0} ^{\infty} \frac{A_{n}}{x^n} ), \quad A_{0} =1
\end{equation}
for an analysis of the asymptotic behaviour. Dividing by $ x^{q+(m-2)/2} $  we have the leading term 
\begin{equation}
\frac{m^2+2m}{4}b +(m+2)bq
\end{equation}
For this to be zero we require  $ q= -m/4$  so that finally we have as $  x>0, x\rightarrow  \infty$
\begin{equation}
\psi \sim exp(-\frac{\sqrt{2}}{m+2} (1+i)|x|^p))|x|^{-m/4} 
\end{equation}
where for $ x>0 $, x is identical with $ | x | $. 
Thus $ \psi \rightarrow 0 $ as $ x \rightarrow \infty$.  In the equivalent  calculation in $ x<0 $ by putting  $ y=-x $ we have the same equation in y except that $ i \rightarrow -i $  so that the analysis is for $ \sqrt{-i}$. An asymptotic solution for $ x<0 $ is therefore
\begin{equation}
\psi \sim exp(-\frac{\sqrt{2}}{m+2} (1-i)|x|^p))|x|^{-m/4} 
\end{equation}
In example considered here,  $ m=1$ and there is a contradiction since if the solution satisfies the asymptotic result, $ \psi_{1} $ cannot satisfy (26) with $ D^2\psi_{1} $ finite at the origin. 
\section{References}
[1] Fring A and  Firth Thomas, Mending the broken PT-regime via an explicit time-dependent Dyson map, Phys. Lett A , 381, 2318,2017 \newline
[2] Fring A, An introduction to PT-symmetric quantum mechanics-time-dependent systems, J.Phys(Conference Series) 2448,012002,2023 \newline
[3]  Fring A,and Moussa M , Unitary quantum evolution for time-dependent quasi-Hermitian systems with nonobservable Hamiltonians, Phsy. Rev A , 93 042114, 2016 \newline
[4] Znojil, M Time-dependent version of crypto-Hermitian quantum theory, Rev.Phys.D, 78 085003, 2008 \newline
[5]Feinberg J and Znojil M,Which Metrics are consistent with a given pseudo-Hermitian matrix, arXiv.2111.04216, 2021\newline
[ 6] Bender C , PT Symmetry in quantum and Classical Physics,World Scientific, 2018 \newline
[7] Burrows B L and Cohen M, Spatial confinement, non-Hermitian  Hamiltonians and related problems, Eur.Phys.D 75,70 (2021) \newline
[8] Mostafazadeh A, Pseudo-Hermiticity versus PT symmetry : The necessary condition for reality of a spectrum of a non-Hermitian Hamiltonian, J. Math. Phys. 43, 205, 2002 \newline
[9] Mostafazadeh A, Pseudo-Hermiticity versus PT symmetry II . A compete characterization of non-Hermitian Hamiltonians with a real spectrum, J.Math. Phys. 43 ,2814, 2002, \newline
[10] Bender C, Making sense of non-Hermitian Hamiltonians, Rep. Prog. Phys. 79 947 2007 \newline
[11] Merzbacher E, Quantum Mechanics,John Wiley and Sons , NewYork 1970

\end{document}